# Structure and photo-induced volume changes of obliquely deposited amorphous selenium

R. Lukács[a.)], J. Hegedüs[b.)], and S. Kugler[a.)]

[a.)]Department of Theoretical Physics, Budapest University of Technology and Economics, H-1521 Budapest, Hungary
[b.)]Department of Chemistry, University of Cambridge, Lensfield Road, Cambridge CB2 1EW, UK

Atomic scale computer simulations on structures and photo induced volume changes of flatly and obliquely deposited amorphous selenium films have been carried out in order to understand how the properties of chalcogenide glasses are influenced by their preparation method. Obliquely deposited a-Se thin films contain more coordination defects, larger voids than the flatly deposited ones. To model the photo induced volume changes the electron excitation and hole creation were treated independently within the framework of tight-binding formalism. Covalent and interchain bond breakings and formations were found. The obliquely deposited samples containing voids showed a wide spectrum of photo induced structural changes in microscopic and volume changes in macroscopic levels.

**INTRODUCTION**

Chalcogenide glasses show a variety of photo induced effects during band-gap illumination like photo-induced volume change (expansion or contraction), photo darkening, photo bleaching, photo densification, etc[1]. It is important to learn their properties because they have technological applications e.g. optical disk, electronic non-volatile memory technology (phase-change random-access memory, PCRAM), etc[2]. Several experiments have been performed since the discovery of photo-induced effects in order to understand their kinetics. One of the first steps was the mapping of structure. At the end of 1970s photo-induced effects were studied by B. Singh et al.[3]. They have concluded that these photo-induced effects in chalcogenide films could be enhanced by oblique deposition. In amorphous GeSe films a correlation was observed between the anomalously large photo contraction and the angle of deposition. Columnar structures were found in these films.

In obliquely deposited amorphous $GeSe_3$ films density decreased when the angle of the evaporant beam increased as it was reported by Rayment and Elliott[4]. Columnar structures of these materials were also found. The illumination with band-gap light caused photo-densification[4]. Columnar structure was observed in obliquely deposited a-$GeS_2$[5], too. Decrease of the refractive index and micro hardness versus angle of incidence has been reported. These are evidences of the increasing free volume with the increase of obliqueness[5]. Photo-induced changes in optical properties of obliquely deposited a-$As_2S_3$ thin films have been studied by Dikova et al.[6]. They found that the increase of refractive index and absorption coefficient is the higher in case of the obliquely deposited films. Recently giant photo-induced expansion was investigated by Ke. Tanaka et al.[7].

A comparison between obliquely deposited As-based and Ge-based chalcogenide films was performed by Shimakawa's group[8]. Photo darkening and photo expansion of As-based chalcogenides and photo bleaching and photo contraction in Ge-based chalcogenides were measured[8]. From these experiments it can be concluded that obliquely deposited chalcogenides show more enhanced photo-induced changes. This could be a consequence of free volume and thus of a more porous structure.

Photo-induced volume expansion in quenched amorphous selenium using tight-binding molecular dynamics (TBMD) computer simulations has been investigated by Hegedüs et al.[9]. They found covalent bond breaking in amorphous networks caused by photo-induced excited electrons, whereas holes contributed to the formation of interchain bonds. Applying bond breaking and interchain bond formation model they described the time development of macroscopic volume expansion in void free amorphous selenium. Recently, Ikeda and Shimakawa[10] published their experimental results on flatly and obliquely deposited a-$As_2Se_3$. Flatly deposited sample shows photo-induced volume expansion while the other shrinks during the illumination and after switching off the light its volume remains the same.

In order to understand these controversial results and the detailed kinetics of the photo-induced changes we have carried out further Molecular Dynamics (MD) simulations on a-Se which is the model material for the chalcogenides. Binary component glass preparation is much more difficult, because at least three different interactions have to be considered[11] during the model construction. As a first stage a set of MD simulations were applied to grow up amorphous samples in different angles of deposition, than the obtained structures of these samples were analyzed[12]. Afterwards a second set of TBMD simulations were carried out to follow up the time development of photoinduced changes on microscopic and macroscopic levels. The structure of this paper is as follows: The simulation method will be presented in the second section, in third and fourth sections the results of the analyses and some conclusions will be formulated in the end.

**METHODS AND MODEL PREPARATION**

A MD computer code[12] had been developed for the simulation of thermal evaporation growth process of flatly and obliquely deposited a-Se. Our purpose was to construct relatively large samples containing at least 1000 atoms. A classical empirical three-body potential was used to calculate the atomic interactions[13]. The simulation technique was the following. A trigonally crystalline lattice, containing 324 selenium atoms was employed to mimic the substrate. There were 108 fixed atoms at the bottom of the substrate and the remaining 216 atoms could move with full dynamics. The simulation cell was open along the positive z axis, and periodic boundary conditions were applied in the x and y directions. Velocity Verlet algorithm was used to follow the atomic motions. The time step was chosen to be 1 fs. Frequency of the atomic injection was f=1/125 fs$^{-1}$ on average. Kinetic energy of the atoms inside the substrate was rescaled at every MD step to keep the substrate at a constant temperature. Several samples were prepared with average angles between the normal to the substrate and the direction of the randomly directed incidence atoms of 0º, 20º, 45º and 60º. Temperature of the substrate was kept at 300 K, while the average bombarding energy was equal to 1 eV.

To simulate the photo induced structural changes in these amorphous samples a second type of MD simulation code were ran using tight-binding (TB) Hamiltonian[14]. This TBMD computer code was developed by Hegedüs et al.[9,15]. It was assumed that immediately after a photon absorption the electron and the hole become separated in space on a femtosecond time scale[16]. Therefore, excited electrons and created holes can be treated independently. Two different types of simulation methods were carried out: first one was to model the excited electron creation: an extra electron was put into the LUMO (Lowest Unoccupied Molecular Orbital), and second case, an electron in the Highest Occupied Molecular Orbital, HOMO (hole creation) was annihilated. In this approach, excitons do not play any role during the photo induced volume changes and the Coulomb interaction between electrons and holes are also neglected.

**STRUCTURE ANALYSIS**

Flatly and obliquely deposited samples have been analyzed in order to obtain information about differences in their structures. Four samples have been considered prepared by incidence average angles of 0º, 20º, 45º and 60º. First the radial distribution functions (RDF) were calculated. Fig.1 shows RDF of the sample deposited under average angle of incidence of 60º.

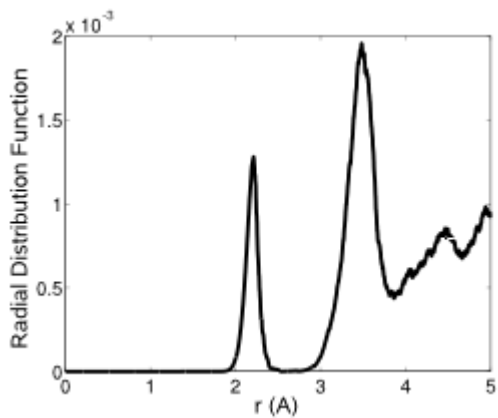

Fig.1. Radial distribution function of an a-Se sample deposited by angle of incidence equal to 60º.

There were no relevant differences among RDFs. The first neighbor shell peak appeared at 2.35 Å, while the second neighbor one was at 3.5 Å. These values provide a good agreement to X-ray diffraction measurements[17], where the pair correlation functions of amorphous Se prepared by mechanical milling (MM) for 50 h and by liquid quenching (LQ) from 600ºC have been derived. The first neighbor distance for the MM-amorphous Se was 2.36 Å and for the LQ-amorphous Se 2.37 Å. The second nearest-neighbor distances were around 3.7 Å for both samples.

Coordination number distributions of the samples were analyzed next. Most of the atoms (>90%) had coordination number of two. There were atoms with coordination number equal to three (~9%) and very few (~1%) with coordination number of one. These are called coordination defects. The number of coordination defects increased by 3% if the average angle of incidence was varied from 0º to 60º.

A correlation was found between the angles of incidence and densities due to different samples i.e. densities decreased monotonically in function of deposition angle in the interval of 0-60 degrees as displayed in Table I.

| Average angle of incidence | Density (g/cm$^3$) |
|---|---|
| 0° | 4.50 |
| 20° | 4.40 |
| 45° | 4.38 |
| 60° | 4.20 |

Table I. Densities of flatly and obliquely deposited samples.

This fact suggests that larger voids could be found in the obliquely deposited films. In order to investigate this supposition, a void size analysis of the samples has been performed by Voronoi-Delaunay approach[18]. The Voronoi diagram of a set of atoms i=1,N is a decomposition of the space into N regions (called Voronoi polyhedra) associated with each atom. i.e. every point of a Voronoi region is closer to the associated atom than to any other atom in the system. Atoms whose Voronoi polyhedras share a face are considered to be contiguous. A set of four atoms contiguous to each other forms a tetrahedron, which is called Delaunay tetrahedron (DT) in three-dimensional space (see two dimensional representation in Fig. 2).

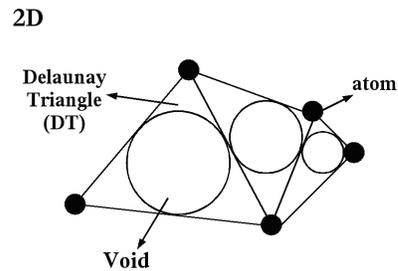

Fig.2. Two-dimensional representation of Delaunay tetrahedrons and a possible approximation of void sizes using a Delaunay-triangle inercircle.

The Delaunay-tetrahedrons insphere volume has been applied as a measure of void volume. Two void size distributions from 3 $Å^3$ to 6 $Å^3$ follow a near logarithmic distribution as shown in Fig. 3.

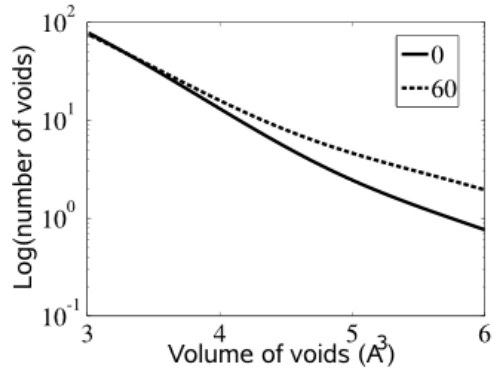

Fig.3. Void size distributions of two different samples in the interval from 3 $Å^3$ to 6 $Å^3$.

Voids with larger volumes than 6 $Å^3$ presented in Fig. 4 appear mostly in the obliquely deposited samples.

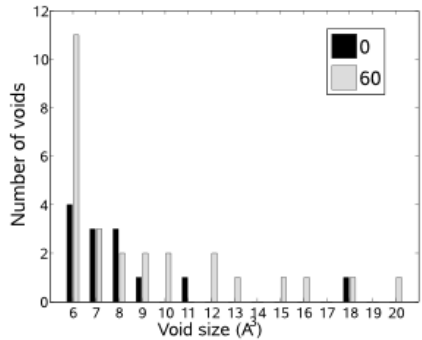

Fig.4. Distributions of largest voids of two different samples.

**PHOTO INDUCED CHANGES**

The structure of amorphous chalcogenide thin films depends strongly on the deposition angle. Our MD computer simulations confirmed that the obliquely deposited films had more porous structures due to the presence of larger voids. A possible scenario for the photo-induced volume contraction of obliquely deposited films proposed by Ikeda and Shimakawa[10] is the following: the primary effect in these materials is the void collapsing. Bond breaking probably destroys voids, decreasing their sizes. The flatly deposited films avoid this effect because of the lack of voids.

TBMD simulations have to be carried out to have an insight in the photo-induced structural changes in amorphous selenium containing voids i.e. the obliquely deposited a-Se film. Thermally well relaxed samples taken from our earlier TBMD simulation[9] containing 162 atoms were used as initial configurations. In the first seria two overlapping ellipsoid voids[4] were artificially created removing Se atoms from the sample in order to make our models to be more realistic. Void creation procedure in this case was the following: the closest atoms to the points at the one third and two third on symmetry line in z direction of the simulation cell were chosen. Centers of the ellipsoids were assigned to these atoms and 16-16 nearest-neighbour atoms to these centers were removed. In the second case several randomly distributed small spherical voids with radius 2.7 Å were made to mimic the obliquely deposited thin films. These new configurations (containing 100 atoms and voids) were relaxed for 20 ps or for 30 ps at 20 K temperature to stabilize the structures. After relaxation either an extra electron (occupation of LUMO increased) or an extra hole (occupation of HOMO decreased) was put into the system. The excited state lasted for 10 ps or for 20 ps. After the excitation processes the samples were relaxed again for 10 ps or for 20 ps.

In void free a-Se films Hegedüs et al. found photo expansion due to covalent bond breaking during the electron excitation process, and photo contraction due to inter-chain bond formation in case of hole creation[9]. Our results of the photo excitation of oblique samples containing voids are presented below.

**A. Electron excitation**

When an extra excited electron was created photo expansion and covalent bond breaking were observed in most of obliquely deposited a-Se films, but some exotic behaviors were also found. The thickness (height) of sample in the open z direction was analysed. Reversible and irreversible photo expansions and contractions were noticed. The reversible photo expansion was induced by a reversible covalent bond breaking as it was derived in our earlier work (See Fig. 3 in Ref. 9). An irreversible covalent bond breaking is displayed in Fig. 5.

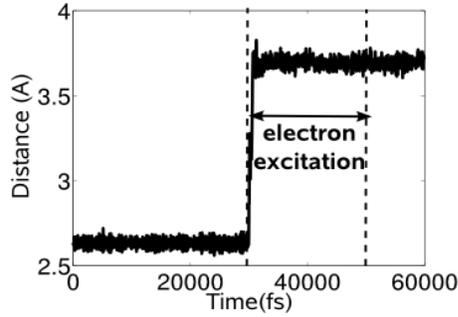

Fig.5. Electron excitation induced irreversible covalent bond breaking.

Time evolution of irreversible volume change is shown in Fig. 6.

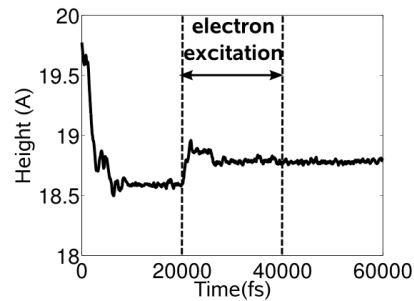

Fig.6. A covalent bond breaking induced irreversible macroscopic photo expansion.

We have found that in such cases a covalent bond breaks irreversibly and causes more irreversible changes in the bonding network (all this is due to the excited electron). In some cases electron excitation can cause reversible photo contraction as shown in Fig. 7.

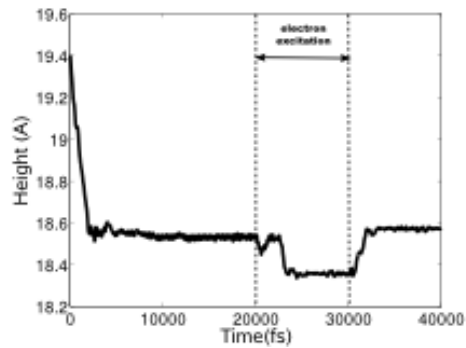

Fig.7. Electron excitation induced reversible macroscopic photo contraction.

The contraction must be caused by the rearrangation of the network because the photo excitation causes reversible bond breaking (elongation) at the same time.

Another exotic case found was a sample showing irreversible photo contraction induced by the electron excitation process (Fig. 8).

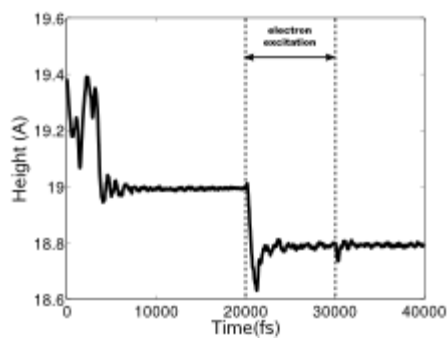

Fig.8. Electron excitation induced irreversible macroscopic photo contraction.

In this case there was an irreversible covalent bond breaking and some irreversible changes in the bonding environment causing void size reduction. During excitation process sample height decreased and further shrinkage occurred after switching off the excitation (this two step process is shown in Fig. 9).

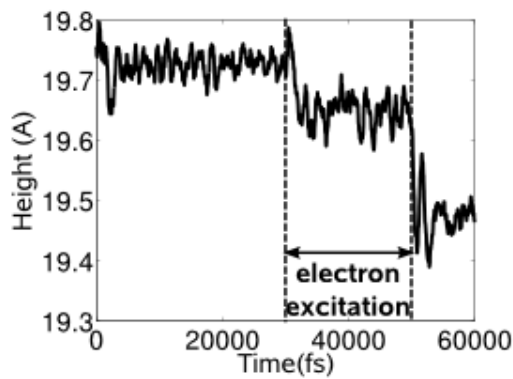

Fig.9. Two step irreversible volume contraction.

Irreversible covalent bond breaking and some irreversible structural changes on the face of the inner voids were found in this case.

**B. Hole creation**

In the second set of TBMD simulations when a hole was created several changes due to the photo-excitation process were observed. More structural changes took place; there were chain deformations, slips, ring creations, covalent/interchain bond formations and breakings. In most cases the height of sample decreased during hole creation and some of the samples only showed transient changes (only present during excitation) due to reversible interchain bond formations (also discussed in Fig.

4 of Ref. 9). However, irreversible covalent (not interchain!) bond formation in the void-containing-network (as shown in Fig. 10) caused irreversible contraction (see Fig. 11).

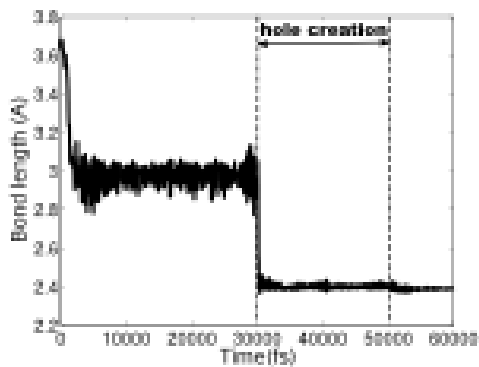

Fig.10. Hole creation induced an irreversible covalent bond formation.

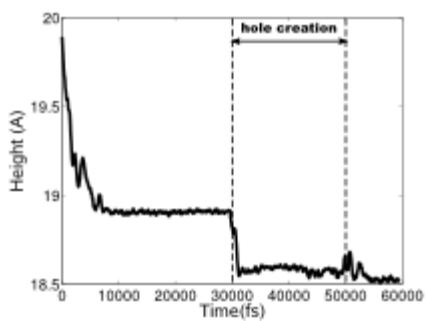

Fig.11. Hole creation induced irreversible macroscopic photo contraction.

In Fig. 12 a composition of reversible and irreversible volume contraction can be seen where the reversible part is small, but significant chain slip was observed inside this sample.

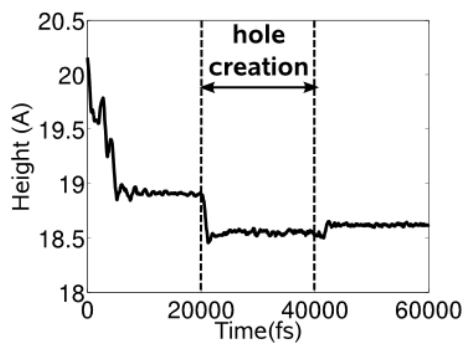

Fig.12. Hole creation induced reversible macroscopic photo contraction.

Some exotic cases were also noticed, where the sample showed reversible (see Fig. 13) or irreversible photo expansion as shown in Fig. 14 but none the less inter-chain bonds were formed.

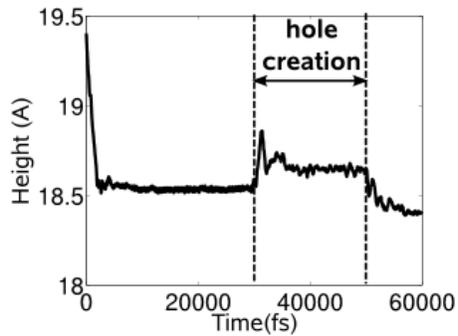

Fig.13. Hole creation induced reversible macroscopic photo expansion.

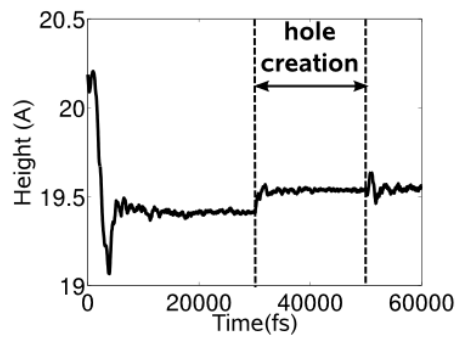

Fig.14. Hole creation induced irreversible macroscopic photo expansion.

**C. Photoinduced bond shift**

In a Se chain a photoinduced covalent bond shift[19] has been observed. Figure 15 shows a snapshot of the Se chain before the illumination.

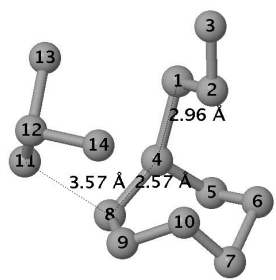

Fig.15. Snapshot of a bonding network before the illumination process.

Covalent bond length was considered to be between 2.2 - 2.6 Å while interchain bonds were between 2.8 − 3.4 Å. The bonding configuration before the illumination presented in Figure 16a was the following: between atom 1 and 4 an interchain bond and between atom 4 and 8 a covalent bond were found. Hole creation process induced some changes in the bonding network. The covalent bond between Se atom 4 and atom 8 got longer and became an interchain bond. Between atom 1 and 4 and atom 8 and 11 covalent bonds were formed (See Fig. 16b.)). After stopping the illumination a covalent bond was formed between atom 1 and 4. Interchain bonds between atom 4 and 8 and between atom 8 and 11 were also formed (Fig. 16c.)). The covalent bond between atom 8 and 4 shifted to atom 4 and 1 as the result of the photo excitation process.

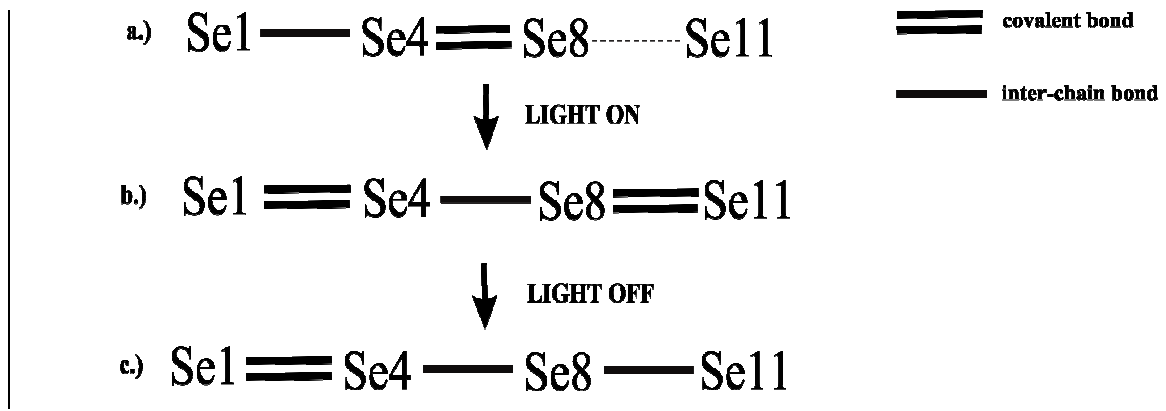

Fig.16. Photoinduced covalent bond shift in an a-Se chain.

In most of the samples structural changes appeared mostly in the vicinity of coordination defects but in few samples chain slips were also found.

**CONCLUSIONS**

Molecular Dynamics simulations have been carried out to investigate structures and photoinduced effects of obliquely deposited a-Se. Increasing deposition angle leads to lower sample density. This rule can be explained by the presence of larger voids which was confirmed by our Voronoi analysis. The TBMD simulations of photoinduced volume changes showed more varieties of microscopic and macroscopic levels than we had obtained in our earlier void free simulations[9]. We described in detail these complex atomic rearrangements occurring on photo-excitation in porous samples which are reversible and irreversible photo contraction/expansion, bond breaking and formations (both inter and intrachain), chain slips and bond shifts. The obliquely deposited chalcogenides are more sensitive to the illumination than the flatly deposited samples because voids provide more degrees of freedom for atomic positions.

ACKNOWLEDGEMENTS

This work has been supported by the OTKA Fund (Grant No. T048699) and Japanese-Hungarian intergovernmental project (No. JP-5/2006). We acknowledge to prof. Koichi Shimakawa (Gifu University, Japan) and to Krisztián Koháry (University of Exeter, UK) for valuable discussions. JH is thankful for the Marie Curie fellowship.